# Spatial variation of Coda *Q* in Kopili fault zone of northeast India as a probe for heterogeneous media

Bora, Nilutpal and Biswas, Rajib

Geophysical Lab, Department of Physics, Tezpur University, Tezpur-784028, Assam
Corresponding author: rajib@tezu.ernet.in

Kopili fault has been experiencing higher seismic and tectonic activity ([1], [2]) during the recent years. These kinds of active tectonics can be inspected by examining coda-wave attenuation and its dependence with frequency. Here, we report spatial variation of coda attenuation of this region. The obtained results reveal that there is velocity anomaly at depth 210-220 km as there arises sharp changes in attenuation coefficient ($\gamma$) and frequency parameter ($n$) which are supported by available data reported by other researchers for this region.

## 1. Introduction

The energy of seismic wave at various distances from earthquake source is severely affected by geological medium. Attenuation is one of the vital parameters that characterizes the medium through which seismic wave propagates. It is defined as the dissipation of seismic energy while passing from the earthquake source to the receiver. This dissipation continues until the seismic wave disappears due to loss of energy (or loss of amplitude). The energy of seismic waves decays due to the geometrical spreading, intrinsic and scattering attenuation [3,4]. The decrease in amplitude of the seismic waves, caused by the geometrical spreading in a homogeneous and isotropic medium, is inversely proportional to the distance travelled. While intrinsic attenuation converts the seismic energy into heat due to inelastic absorption, scattering or elastic attenuation redistributes the energy at random heterogeneities present in the earth's crust. It should be pointed out that the total energy in the wavefield remains constant in case of scattering attenuation, whereas the intrinsic attenuation causes disappearance of the wave due to loss of energy. Therefore, attenuation of seismic waves in the earth's crust is a key phenomenon for understanding the diversity in the earth's physical states and seismic potential of the region [5,6]. The attenuation property of the medium is usually measured by a dimensionless quantity, called the quality factor Q, which is defined as the ratio of wave energy to the energy lost per harmonic oscillations [7,8] while propagating through the intermediate medium. The attenuation of medium is inversely proportional to the quality factor (Q) which means that, seismic waves are highly attenuated for regions having lower Q values.

**FIGURE 1 HERE**

**Caption:** (a) Simplified map of Northeast India. (b) Map showing major tectonic features of Kopili Fault region. These are designated as: Main Boundary Thrust (MBT), Main Central Thrust (MCT), Dauki fault, Kopili fault (KF), Oldham fault (OF), Borapani Shear Zone (BS). The filled circles represent the epicenters used for our study and triangles represents the stations temporarily deployed in Kopili region. The star symbols represent the Great and major earthquakes originating in and around Shillong Plateau.

There are many literatures available regarding the estimation of $Q_C$ [9-21]. Adding to this, Aki (1969), Aki and Chouet (1975) and Sato (1977) suggested that by analyzing tail portion of earthquakes waveform data permits us to study the wave attenuation mechanism in the earth's interior. The Kopili fault extends from western part of Manipur to tri-junction of Bhutan, Arunachal Pradesh and Assam. It's length is roughly 300km and width is 50km. Very recently [22] cited about the role of Kopili fault in the recent Imphal earthquake. Furthermore, [1] estimates the body wave attenuation mechanism for this study area using direct *P* and *S*-wave. On the other hand, [23] investigated the coda wave attenuation mechanism and opinioned that there was some velocity anomaly underneath this region. Our this study will mainly focused on the anomaly obtained by [23] and regarding this we also try to developed a simplified 3D profile of the region. While addressing these anomaly, we try to correlate them with variation in velocity and temperature as a probable reason to apprehend the inherent mechanism.

## 2. Dataset and Methodology

The present dataset involves *300* digital seismograms of several earthquakes having magnitude $M_L$= *2.1-3.9*, recorded by six stations (Figure 1). This network of stations is operated by National Geophysical Research Institute (NGRI), Hyderabad. All the six stations are equipped with three component Broadband Seismographs with GPS synchronized timings. The data were sampled with a digitizing frequency of *100* samples per second. The corresponding epicentral plot is provided in Figure 1. As seen, the events are randomly scattered around the receiver sites under consideration.

We have adopted the single backscattering model of [7] to measure seismic wave attenuation from the coda wave. As we have utilized the same dataset as that of [23], therefore we followed there paper in order to estimate the $Q_C$.

In order to estimate the depth variation, as defined by [7, 24-26] the estimated $Q_C$ represents the normal decay of amplitude of the back-scattered wave for an ellipsoidal volume where the earthquake source and receiver are located at its focus. Therefore, the surface area of this ellipsoidal volume inside which the coda waves are distributed can be represented using the following equation:

$$\frac{x^2}{a^2} + \frac{y^2}{b^2} = 1 \tag{1}$$

where, $a = vt/2$ and $b = \sqrt{a^2 - \frac{\Delta^2}{4}}$. Here, *v*, $\Delta$ and *t* represents the s-wave velocity, average hypocentral distance and average length of the lapse time. In general, the average lapse time is taken as $t = t_0 + \frac{W}{2}$ (where $t_0$ stands for starting time of coda window and w is the coda window length). It is seen that if the value of $\Delta$ equals to zero than the above equation represent a circular area of radius $vt/2$. The lower border of this ellipsoid or the maximum depth of this assumed ellipsoid represent the penetration depth of the estimated coda wave quality factor [27] and can be calculated by $h = h_{avg} + b$ in which $h_{avg}$ is the average earthquake depth.

### 3. Data Analysis

As reported by [3] and [28], there is slight difference between vertical and horizontal components of the S-coda envelopes of small earthquake. For simplicity, we have only used the vertical components for our study. Each seismogram was bandpass-filtered at five frequency bands (1-2; 2-5; 4-8; 6-12; 9-15 Hz) with central frequencies at 1.5, 3.5, 6, 9 and 12 Hz, by using a fourth-order Butterworth filter respectively. Then, the coda amplitudes $A_C(f,t)$ were calculated as per [23] by following the simplified coda back scattering method.

### 4. Results and Discussion

As reported by [23], for window length of 30sec, the average value of $Q_C$ increases from 114 at frequency 1.5Hz to 1563 at frequency 12Hz. Similarly, an increasing trend in $Q_C$ was observed for all the lapse time window up to 90sec. A rough estimate of maximum depth through lapse time and average epicentral distance is achieved using the equation 1. Here, we have taken $v = 3.5$ km/s. The value of $w$ is taken from 30 to 90 sec. As established by [23], the obtained values of the axis are almost equal, therefore we can considered the ellipsoidal volume practically spherical.

In this chapter, the $Q_0$ and n values are taken from the results obtained by [23], as we have utilized the same dataset here.

#### 4.1. Lateral and Spatial variation of Coda Q

In this phase, we attempt to estimate the 3D velocity model in order to understand the underneath subsurface structure. We applied the results obtained from [23], checking different gridding parameters to provide a reliable and robust 3D model without sacrificing its resolution.

Different parameterization schemes also were applied. Based on the geometry of the survey points and assuming the near-surface velocity and thickness to be the important factors, the optimum result was derived using a 2x2x1-km (along the x-, y-, and z-axes) grid spacing on a linear interpolation scheme with values every 1 km. Although efforts were made to minimize the grid spacing further, the survey points and the depth profiles estimated in this study were distributed unevenly; the resulting model showed undesirable velocities. In our research, a 2x2x1-km grid spacing was eventually selected for interpretation. We used the ordinary kriging estimation procedure as an interpolation scheme [29]. The ordinary kriging interpolation method allows us to characterize an unknown regionalized variable (a spatially continuous, random function with some geographical distribution) from the samples in a neighborhood of any unsampled location [30].

**FIGURE 2 HERE**

**Caption:** (a) The study region and the dotted box represents the area selected for the 3D profile. Triangles are stations used to draw maps. (b) and (c) Shows the cross section along depth for $Q_0$ and n values. The scale indicates the maximum and minimum values of $Q_0$ and n. The Solid black arrow shows the north direction.

In order to carry out this we have chosen identical epicentral range in each data set. This is essential because longer rays penetrate to greater depths, where attenuation might be different, and two stations with similar attenuation properties around them could show different $Q_0$ if one receives rays from farther distances. In order to have similar epicentral distance we have selected only those so that the average epicentral distance remain in the range of 80 to 90km. A rough estimate of maximum depth through lapse time and average epicentral distance is achieved using the equation 1. Here, we have taken $v = 3.5$ km/s [31]. The value of *w* is taken from 30 to 90 sec. The observed variations are shown in the figure 2 and 3.

As seen in Figure 2(b), the $Q_0$ values have lower values in the northern side of the region which are indicated by light and dark blue colors. This means, that along this side, the seismic waves are more attenuated. But, as we go deeper there we noticed a change in color which is linked with the heterogeneity as mentioned by [23]. Similarly, the station DMK shows some variation in the depth range 230-270km. This is also linked with the underlying geology. In order to further investigate this, we have found a noticeable evidence from [2], as they reported that as DMK is located on an alluvial plain, which generally traps seismic waves. As a result, the waves are attenuated. But with increase in coda window length, the values of $Q_0$ increases which may be due to the higher penetration depth since the lithosphere becomes more homogeneous with depth.

**FIGURE 3 HERE**

**Caption:** (a) The study region and the dotted box represents the area selected for the 3D profile. Triangles are stations used to draw maps. (b) to (e) Shows the 3D profile along with the iso-surfaces. The values of $Q_0$ for all the iso-surfaces are given at the top of each figure. The scale indicates the maximum and minimum values of $Q_0$. The Solid black arrow shows the north direction.

The stations BPG, RUP and BKD show high n values which are linked with their seismic profile as all these stations are situated above MBT. As, higher n values show that these sites are more heterogeneous than the remaining site. Similar pattern is also visible in $Q_0$ values [figure 2(c)]. These lower $Q_0$ values reflect strong scattering effects arising either due to the dense faulting or complex structure beneath these three stations. As it is well known that lesser the value of $Q_0$, higher seismic activity is prevalent. The reason behind this may be attributed to the presence of MBT (main boundary thrust) near all these three stations, which makes them highly active as higher values of *n* indicate active region. It is worthy to mention here that MBT is one of the main tectonic features of Himalayas region and most earthquakes that occur in the Himalayas are along these tectonic features. It is claimed that the Kopili fault intersects the Himalaya and caused displacement and curvilinear structure at the MBT and MCT (Main Central Thrust) zone [32, 33]. Since it is well known that the Kopili fault segmented the Shillong Platue into two parts namely, the Shillong Massif and the Mikir Massif.

But when we go deeper, we have found a sharp decrease in *n* values which also indicate some anomaly present there. In the south part of the region, we also seen some anomaly. As it is clearly indicated by the change in color.

All these variations are smoothly represented in figure 3 with the help of iso-surface. As indicated in figure 3(b), there are almost same values ($Q_0$=110) in the upper and lower part of the profile. But as we increase the values, we have seen that the top and bottom iso-surface profile are moves towards each other. That is to say, the values of $Q_0$ increases toward the mid layer portion of the study area. In figure 3(e), we have seen at $Q_0$=185 (which is one of the highest values $Q_0$), both the upper and lower iso-surfaces collide. That means, at these portion, the attenuation is less and also represent one almost solid layer as we can see a thickness in between the iso-surfaces. Which indicates a strong anomaly in the values of $Q_0$ distributed in this layer. It can be proposed that there may be presence of rock to explain the high velocity/low attenuation zone beneath the lithosphere.

Another plausible reason for this variation may be ascribed to either prevalence of strong heterogeneity in the lithospheric mantle or possibly significant variations of rock composition and temperature in the uppermost mantle. This conjecture gets further validation from the report of [34] where they hinted at significant variation of FVD (Fast velocity direction) in the Indian lithosphere in the depth range of 60-200 km, implicating anisotropy. [35] gave a hypothesis in order to explain presence of seismic discontinuities in the depth range over 250km, invoking interplay of an implicit low velocity zone. They [35] reported that another possibility of this discontinuity may involve the anomalous nature of melt migration and segregation at high pressure and temperatures corresponding to depth as deep as 180-300km.

As it was already established that anomalies of seismic velocity and attenuation are correlated. Various studies demonstrated that high-Q zones in the upper mantle are well correlated with high-velocity zones beneath active volcanoes where partial melting is expected. A similarity between the observed relationship and predictions based on experimental studies [36], permitted the authors to conclude that both attenuation and velocity anomalies should be primarily of thermal origin [37].

Among many other physical phenomena that affect $Q$, temperature also probably plays a dominant role below the upper crust [38]. It is widely accepted that the $Q_C$ value in high temperature or magma intrusion area is usually low. [38] have shown that the $Q^{-1}$ variations are a reasonable proxy for temperature variations. Gao [39] suggested that, if coda mainly reveals the intrinsic $Q$, it should be very sensitive to temperature and the content of liquid within the detected body. Conversely, if the activities are produced by small stress variations, consideration of coda $Q$ may be less sensitive to the change in stress in comparison to temperature; it may thus be less sensitive to the stress related precursors. The data of geothermal structure for the crust and mantle indicate that the heat flow for the region is as high as 51-70 mW/m$^2$ (data from International Heat Flow Commission). As reported by [40], that the distribution of earthquake depths throughout this region is consistent with a generic global view of seismicity in which earthquakes occur in (1) 'wet' upper crustal material to a temperature of ~350$^0$C, or (2) higher temperatures in dry granulite-facies lower crust or (3) mantle that is colder than ~600$^0$C. The physical mechanism that $Q_0$ usually exhibits low values in the tectonically active regions still needs more in-depth analysis. Mitchell [41] pointed out that tectonic activity generates heat and the heat initiates hydrothermal reactions in the upper mantle and, perhaps, in the crust. The fluids, caused by the hydrothermal reactions after release, flow upward through cracks and permeable rocks. Sometimes, they predominantly habitat in the upper crust. As such, regions of low *Q* can be, in general, characterized by high temperatures and heat flow. Further, the cracks filled with liquids, as aforementioned enhances the scattering, resulting in loss of the energy of the waves. This might be the reason of low *Q* in the tectonically active regions. The data of geothermal structure for the crust and mantle indicate that the heat flow for the region is as high as 51-70 mW/m$^2$ (data from International Heat Flow Commission). Thus the low coda *$Q_0$* value obtained in the present study is supported by the heat flow and high temperature.

## 5. Conclusions:

In summary, we report spatial variation of coda Q along Kopili fault. With considerable evidence of attenuation anomalies at certain depth, the findings are correlated with seismic activity, velocity anomalies as well as temperature. We have found high $Q_0$ at a depth 230-270 km, which has been well linked with high velocity zones. The attained results bear a good signature with prevalent mechanism of underneath medium. It is expected that these findings will pave way for deeper geophysical investigations that will help giving insight towards understanding underneath attenuation mechanism.

[41] Mitchell, B. J. Anelastic Structure and Evolution of the Continental Crust and Upper Mantle from Seismic Surface Wave Attenuation. Reviews of Geophysics. 33(4) (1995) 441.

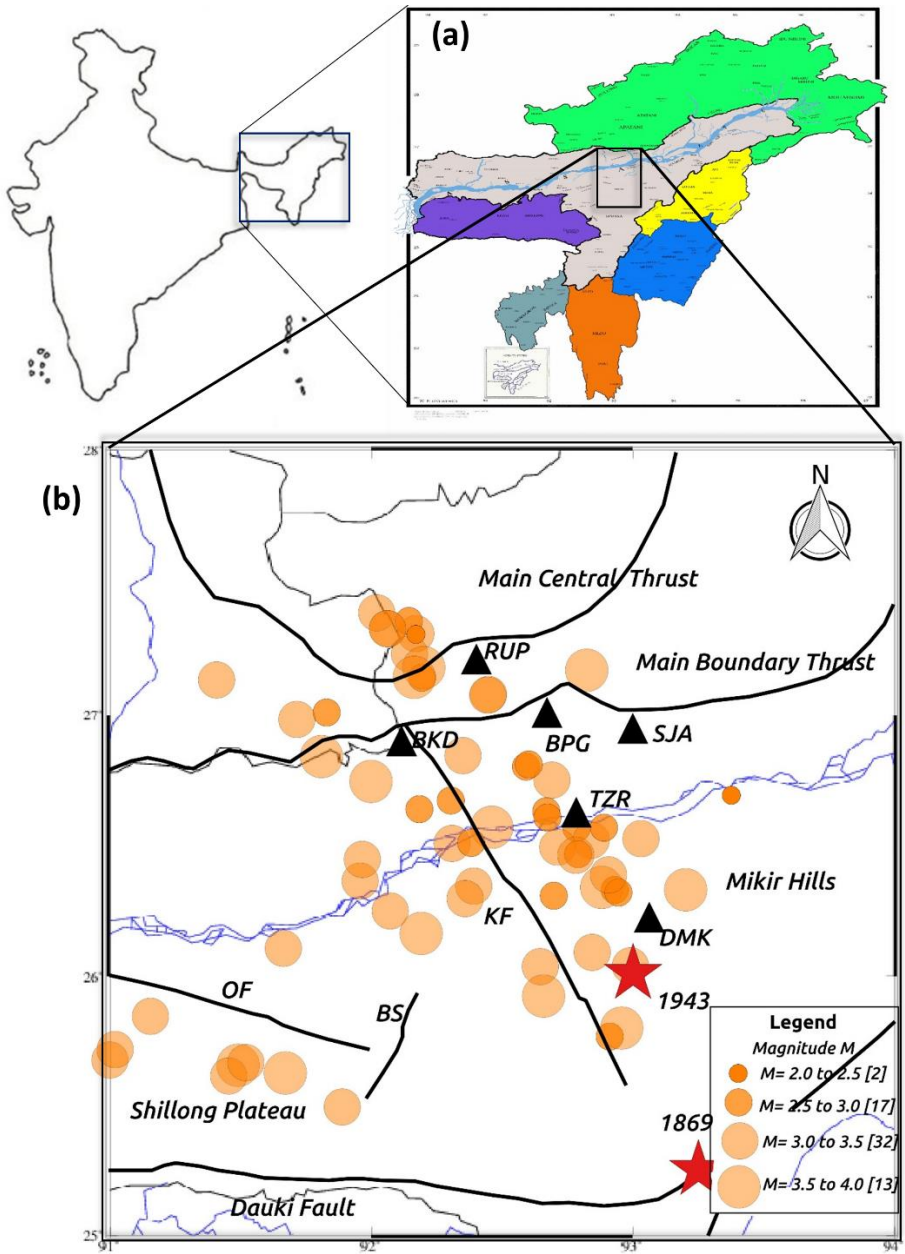

**Figure 1:** (a) Simplified map of Northeast India. (b) Map showing major tectonic features of Kopili

Fault region. These are designated as: Main Boundary Thrust (MBT), Main Central Thrust (MCT), Dauki fault, Kopili fault (KF), Oldham fault (OF), Borapani Shear Zone (BS). The filled circles represent the epicenters used for our study and triangles represents the stations temporarily deployed in Kopili region. The star symbols represent the Great and major earthquakes originating in and around Shillong Plateau.

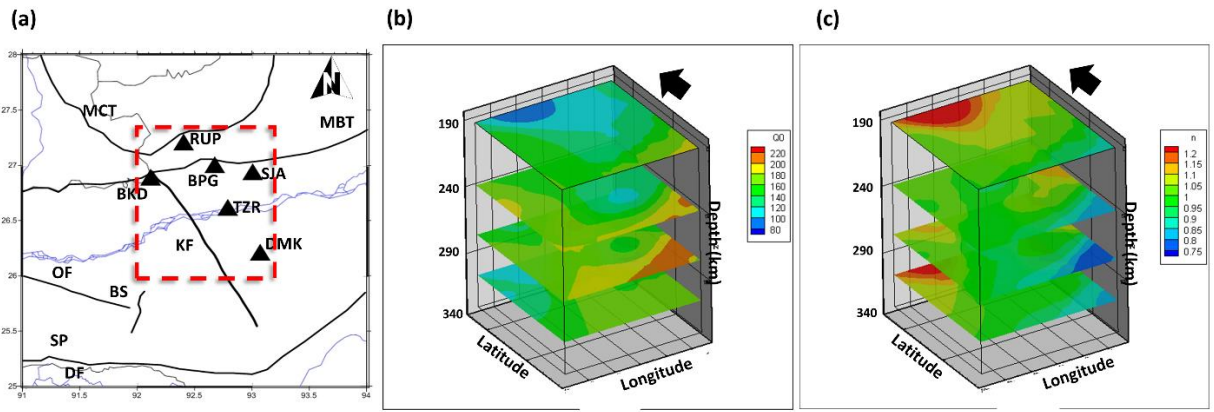

**Figure 2:** (a) The study region and the dotted box represents the area selected for the 3D profile. Triangles are stations used to draw maps. (b) to (e) Shows the 3D profile along with the iso-surfaces. The values of $Q_0$ for all the iso-surfaces are given at the top of each figure. The scale indicates the maximum and minimum values of $Q_0$. The Solid black arrow shows the north direction.

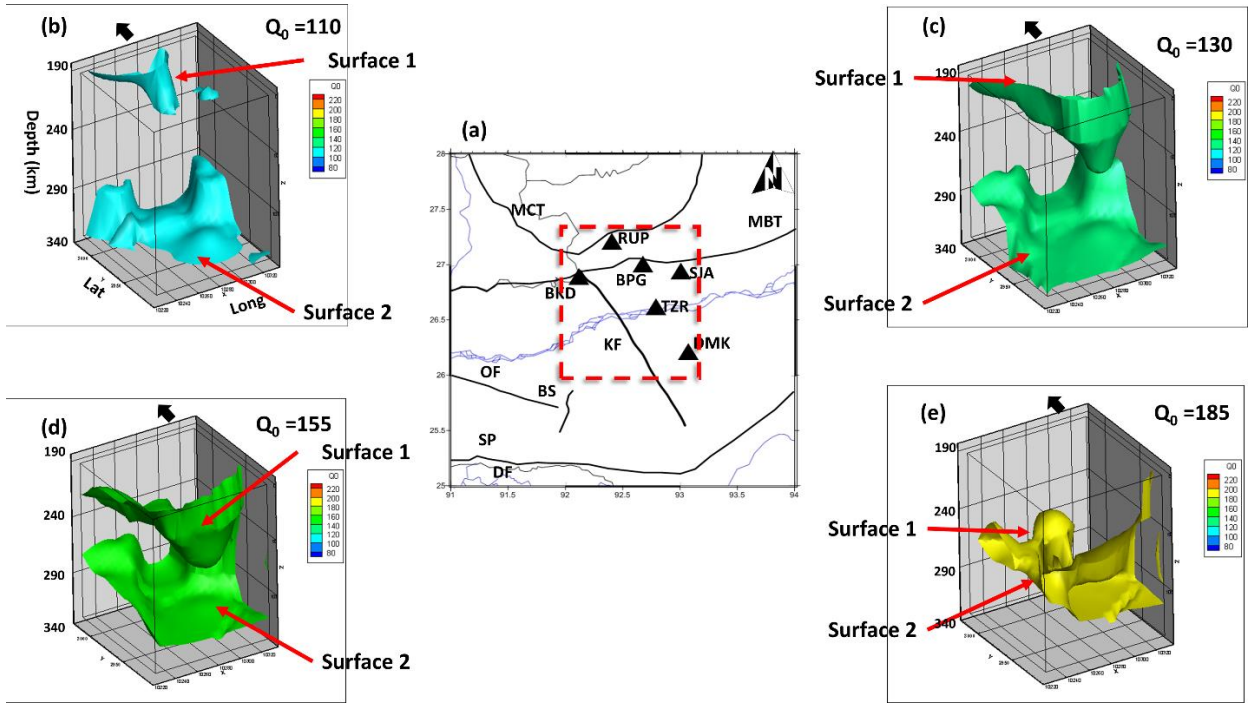

**Figure 3:** (a) The study region and the dotted box represents the area selected for the 3D profile. Triangles are stations used to draw maps. (b) to (e) Shows the 3D profile along with the iso-surfaces. The values of $Q_0$ for all the iso-surfaces are given at the top of each figure. The scale indicates the maximum and minimum values of $Q_0$. The Solid black arrow shows the north direction.